\input harvmac

\Title{\vbox{\baselineskip12pt
\hbox{WIS/19/02-MAY-DPP}
\hbox{hep-th/0205155}
}}
{\vbox{\centerline{Duality cascades}
\centerline {and duality walls.}}}

\centerline{Bartomeu Fiol}

\bigskip
\medskip

{\vbox{\centerline{ \sl Department of Particle Physics}
 \centerline{\sl The Weizmann Institute of Science}
\vskip2pt
\centerline{\sl Rehovot, 76100, Israel}}
\centerline{ \it fiol@clever.weizmann.ac.il }}

\bigskip
\bigskip
\noindent

\vskip .1in\centerline{\bf Abstract}
We recast the phenomenon of duality cascades in terms of the
Cartan matrix associated to the quiver gauge theories appearing
in the cascade. In this language, Seiberg dualities for the different
gauge factors correspond to Weyl reflections. We argue that the UV
behavior of different duality cascades depends markedly on whether
the Cartan matrix is affine ADE or not. In particular, we find examples
of duality cascades that can't be continued after a finite energy scale,
reaching a ``duality wall'', in terminology due to M. Strassler. For these 
duality cascades we suggest the existence of a UV completion in 
terms of a little string theory.

\let\includefigures=\iftrue
%
\let\useblackboard=\iftrue
%
%
\newfam\black
\input epsf
\includefigures
\message{If you do not have epsf.tex (to include figures),}
\message{change the option at the top of the tex file.}
\def\figin{\epsfcheck\figin}\def\figins{\epsfcheck\figins}
\def\epsfcheck{\ifx\epsfbox\UnDeFiNeD
\message{(NO epsf.tex, FIGURES WILL BE IGNORED)}
\gdef\figin##1{\vskip2in}\gdef\figins##1{\hskip.5in}
\else\message{(FIGURES WILL BE INCLUDED)}%
\gdef\figin##1{##1}\gdef\figins##1{##1}\fi}
\def\DefWarn#1{}
\def\figinsert{\goodbreak\midinsert}
\def\ifig#1#2#3{\DefWarn#1\xdef#1{fig.~\the\figno}
\writedef{#1\leftbracket fig.\noexpand~\the\figno}%
\figinsert\figin{\centerline{#3}}\medskip\centerline{\vbox{\baselineskip12pt
\advance\hsize by -1truein\noindent\footnotefont{\bf Fig.~\the\figno:} #2}}
\bigskip\endinsert\global\advance\figno by1}
\else
\def\ifig#1#2#3{\xdef#1{fig.~\the\figno}
\writedef{#1\leftbracket fig.\noexpand~\the\figno}%
\global\advance\figno by1}
\fi

\def\IZ{\relax\ifmmode\mathchoice
{\hbox{\cmss Z\kern-.4em Z}}{\hbox{\cmss Z\kern-.4em Z}}
{\lower.9pt\hbox{\cmsss Z\kern-.4em Z}}
{\lower1.2pt\hbox{\cmsss Z\kern-.4em Z}}\else{\cmss Z\kern-.4em
Z}\fi}
\def\inbar{\,\vrule height1.5ex width.4pt depth0pt}
\def\IB{\relax{\rm I\kern-.18em B}}
\def\IC{\relax\hbox{$\inbar\kern-.3em{\rm C}$}}
\def\ID{\relax{\rm I\kern-.18em D}}
\def\IE{\relax{\rm I\kern-.18em E}}
\def\IF{\relax{\rm I\kern-.18em F}}
\def\IG{\relax\hbox{$\inbar\kern-.3em{\rm G}$}}
\def\IH{\relax{\rm I\kern-.18em H}}
\def\II{\relax{\rm I\kern-.18em I}}
\def\IK{\relax{\rm I\kern-.18em K}}
\def\IL{\relax{\rm I\kern-.18em L}}
\def\IM{\relax{\rm I\kern-.18em M}}
\def\IN{\relax{\rm I\kern-.18em N}}
\def\IO{\relax\hbox{$\inbar\kern-.3em{\rm O}$}}
\def\IP{\relax{\rm I\kern-.18em P}}
\def\IQ{\relax\hbox{$\inbar\kern-.3em{\rm Q}$}}
\def\IR{\relax{\rm I\kern-.18em R}}
\font\cmss=cmss10 \font\cmsss=cmss10 at 7pt
\def\IZ{\relax\ifmmode\mathchoice
{\hbox{\cmss Z\kern-.4em Z}}{\hbox{\cmss Z\kern-.4em Z}}
{\lower.9pt\hbox{\cmsss Z\kern-.4em Z}}
{\lower1.2pt\hbox{\cmsss Z\kern-.4em Z}}\else{\cmss Z\kern-.4em Z}\fi}
\def\IGa{\relax\hbox{${\rm I}\kern-.18em\Gamma$}}
\def\IPi{\relax\hbox{${\rm I}\kern-.18em\Pi$}}
\def\ITh{\relax\hbox{$\inbar\kern-.3em\Theta$}}
\def\IOm{\relax\hbox{$\inbar\kern-3.00pt\Omega$}}

\def\inbar{\,\vrule height1.5ex width.4pt depth0pt}

\font\cmss=cmss10 \font\cmsss=cmss10 at 7pt
\def\IR{\relax{\rm I\kern-.18em R}}

\lref\fm{B. Fiol and M. Mari\~no, ``BPS states and algebras from quivers'',
JHEP {\bf 0007} (2000), 31.}

\lref\tata{
K.~h.~Oh and R.~Tatar,
arXiv:hep-th/0112040.
}

\lref\ttata{
K.~Dasgupta, K.~Oh and R.~Tatar,
arXiv:hep-th/0106040.
}

\lref\tttata{
K.~Oh and R.~Tatar,
JHEP {\bf 0005}, 030 (2000)
[arXiv:hep-th/0003183].
}

\lref\spcasca{
S.~G.~Naculich, H.~J.~Schnitzer and N.~Wyllard,
``A cascading N = 1 Sp(2N+2M) x Sp(2N) gauge theory,''
arXiv:hep-th/0204023.
}

\lref\imai{
S.~Imai and T.~Yokono,
Phys.\ Rev.\ D {\bf 65} 066007 (2002)
[arXiv:hep-th/0110209].
}

\lref\ahn{
C.~h.~Ahn, S.~k.~Nam and S.~J.~Sin,
Phys.\ Lett.\ B {bf 517}, 397 (2001)
[arXiv:hep-th/0106093].
}

\lref\is{
K.~A.~Intriligator and N.~Seiberg,
Nucl.\ Phys.\ B {\bf 444}, 125 (1995)
[arXiv:hep-th/9503179].
}

\lref\dfr{
M.~R.~Douglas, B.~Fiol and C.~Romelsberger,
``The spectrum of BPS branes on a noncompact Calabi-Yau,''
arXiv:hep-th/0003263.
}

\lref\dgm{ 
M.~R.~Douglas, B.~R.~Greene and D.~R.~Morrison,
Nucl.\ Phys.\ B {\bf 506}, 84 (1997)
[arXiv:hep-th/9704151].
}

\lref\dm{
M.~R.~Douglas and G.~W.~Moore,
``D-branes, Quivers, and ALE Instantons,''
arXiv:hep-th/9603167.
}

\lref\kac{V.G. Kac, ``Infinite root systems, representations of 
graphs and invariant theory,'' Inv. Math. {\bf 56} (1980) 57.}

\lref\kn{
I.~R.~Klebanov and N.~A.~Nekrasov,
Nucl.\ Phys.\ B {\bf 574}, 263 (2000)
[arXiv:hep-th/9911096].
}

\lref\ks{
I.~R.~Klebanov and M.~J.~Strassler,
JHEP {\bf 0008}, 052 (2000)
[arXiv:hep-th/0007191].
}

\lref\kw{
I.~R.~Klebanov and E.~Witten,
Nucl.\ Phys.\ B {\bf 536}, 199 (1998)
[arXiv:hep-th/9807080].
}

\lref\cfikv{
F.~Cachazo, B.~Fiol, K.~A.~Intriligator, S.~Katz and C.~Vafa,
``A geometric unification of dualities,''
arXiv:hep-th/0110028.
}

\lref\gns{
S.~Gubser, N.~Nekrasov and S.~Shatashvili,
JHEP {\bf 9905}, 003 (1999) [hep-th/9811230].
}

\lref\ofer{
O.~Aharony,
JHEP {\bf 0103}, 012 (2001)
[arXiv:hep-th/0101013].
}

\lref\seiberg{
N.~Seiberg,
Nucl.\ Phys.\ B {\bf 435}, 129 (1995)
[arXiv:hep-th/9411149].
}

\lref\sw{
N.~Seiberg and E.~Witten,
Nucl.\ Phys.\ B {\bf 426}, 19 (1994)
[Erratum-ibid.\ B {\bf 430}, 485 (1994)]
[arXiv:hep-th/9407087].
}

\lref\lnv{
A.~E.~Lawrence, N.~Nekrasov and C.~Vafa,
Nucl.\ Phys.\ B {\bf 533}, 199 (1998)
[arXiv:hep-th/9803015].
}

\lref\lopez{
E.~Lopez,
JHEP {\bf 9902}, 019 (1999)
[arXiv:hep-th/9812025].
}

\lref\kmv{
S.~Katz, P.~Mayr and C.~Vafa,
Adv.\ Theor.\ Math.\ Phys.\  {\bf 1}, 53 (1998)
[arXiv:hep-th/9706110].
}

\lref\polchi{
M.~Bertolini, P.~Di Vecchia, M.~Frau, A.~Lerda, R.~Marotta and I.~Pesando,
JHEP {\bf 0102}, 014 (2001)
[arXiv:hep-th/0011077].
 J.~Polchinski,
Int.\ J.\ Mod.\ Phys.\ A {\bf 16}, 707 (2001)
[arXiv:hep-th/0011193].
}

\lref\egkrs{
S.~Elitzur, A.~Giveon, D.~Kutasov, E.~Rabinovici and A.~Schwimmer,
Nucl.\ Phys.\ B {\bf 505}, 202 (1997)
[arXiv:hep-th/9704104].
}

\lref\witten{
E.~Witten,
Nucl.\ Phys.\ B {\bf 500}, 3 (1997)
[arXiv:hep-th/9703166].
}

\lref\gk{
A.~Giveon, D.~Kutasov and O.~Pelc,
JHEP {\bf 9910}, 035 (1999)
[arXiv:hep-th/9907178].
}


\lref\italians{
M.~Bertolini, P.~Di Vecchia and R.~Marotta,
``N = 2 four-dimensional gauge theories from fractional branes,''
arXiv:hep-th/0112195.
}

\lref\fkt{
S.~Frolov, I.~R.~Klebanov and A.~A.~Tseytlin,
Nucl.\ Phys.\ B {\bf 620}, 84 (2002)
[arXiv:hep-th/0108106].
}

\lref\ls{
R.~G.~Leigh and M.~J.~Strassler,
Nucl.\ Phys.\ B {\bf 447}, 95 (1995)
[arXiv:hep-th/9503121].
}

\lref\kt{
I.~R.~Klebanov and A.~A.~Tseytlin,
Nucl.\ Phys.\ B {\bf 578}, 123 (2000)
[arXiv:hep-th/0002159].
}
        
\lref\gubke{
S.~S.~Gubser and I.~R.~Klebanov,
Phys.\ Rev.\ D {\bf 58}, 125025 (1998)
[arXiv:hep-th/9808075].
}

\lref\strassler{M. J. Strassler, ``Duality in Supersymmetric Field
Theory and an Application to Real Particle Physics''. Talk given
at International Workshop on Perspectives of Strong Coupling Gauge 
Theories (SCGT 96), Nagoya, Japan. Available at
http://www.eken.phys.nagoya-u.ac.jp/Scgt/proc/}

\lref\adsrev{
O.~Aharony, S.~S.~Gubser, J.~Maldacena, H.~Ooguri and Y.~Oz,
Phys.\ Rept.\  {\bf 323}, 183 (2000)
[arXiv:hep-th/9905111].
}

\lref\vafa{
C.~Vafa,
``Mirror symmetry and closed string tachyon condensation,''
arXiv:hep-th/0111051.
}

\lref\strass{
M.~J.~Strassler,
``On methods for extracting exact non-perturbative results in  
non-supersymmetric gauge theories,''
arXiv:hep-th/0104032.
}

\lref\schmaltz{
M.~Schmaltz,
Phys.\ Rev.\ D {\bf 59}, 105018 (1999)
[arXiv:hep-th/9805218].
}

\lref\ab{
O.~Aharony, and M.~Berkooz,
JHEP {\bf 9910}, 030 (1999)
[arXiv:hep-th/9909101].
}

\lref\abks{
O.~Aharony, M.~Berkooz, D.~Kutasov and N.~Seiberg,
JHEP {\bf 9810}, 004 (1998)
[arXiv:hep-th/9808149].
}

\lref\uranga{
A.~M.~Uranga,
JHEP {\bf 9901}, 022 (1999)
[arXiv:hep-th/9811004].
}

\lref\exact{
N.~Seiberg,
Phys.\ Rev.\ D {\bf 49}, 6857 (1994)
[arXiv:hep-th/9402044].
}

\lref\itzhaki{
N.~Itzhaki, J.~M.~Maldacena, J.~Sonnenschein and S.~Yankielowicz,
Phys.\ Rev.\ D {\bf 58}, 046004 (1998)
[arXiv:hep-th/9802042].
}

\lref\peet{Peet, Polchinski}

\lref\hko{
C.~P.~Herzog, I.~R.~Klebanov and P.~Ouyang  [C01-07-20.4 Collaboration],
``Remarks on the warped deformed conifold,''
arXiv:hep-th/0108101.
}

\lref\yanghui{
Y.~H.~He,
``Some remarks on the finitude of quiver theories,''
arXiv:hep-th/9911114.
}

\lref\oovafa{
H.~Ooguri and C.~Vafa,
Nucl.\ Phys.\ B {\bf 500}, 62 (1997)
[arXiv:hep-th/9702180].
}

\lref\feng{
B.~Feng, A.~Hanany, Y.~H.~He and A.~M.~Uranga,
JHEP {\bf 0112}, 035 (2001)
[arXiv:hep-th/0109063].
}

\lref\beas{
C.~E.~Beasley and M.~R.~Plesser,
JHEP {\bf 0112}, 001 (2001)
[arXiv:hep-th/0109053].
}

\lref\brax{
P.~Brax, A.~Falkowski, Z.~Lalak and S.~Pokorsky,
``Custodial supersymmetry in non-supersymmetric 
quiver theories,''
arXiv:hep-th/0204195.
}

\lref\gimon{
E.~G.~Gimon and J.~Polchinski,
Phys.\ Rev.\ D{\bf 54}, 1667 (1996)
[arXiv:hep-th/9601038].
}

\lref\revtwo{
C.~P.~Herzog, I.~R.~Klebanov and P.~Ouyang,
``D-branes on the Conifold and N=1 Gauge/Gravity Dualities''
arXiv:hep-th/0205100.
}

\lref\bertol{
M.~Bertolini, P.~Di Vecchia, M.~Frau, A.~Lerda, R.~Marotta and I.~Pesando,
JHEP {\bf 0102}, 014 (2001)
[arXiv:hep-th/0011077]
}

\lref\malda{
J.~M.~Maldacena and A.~Strominger,
JHEP {\bf 9712}, 008 (1997)
[arXiv:hep-th/9710014].
}

\Date{May 2002}

\newsec {Introduction.}

One of the most interesting developments in the AdS/CFT correspondence 
\adsrev\ has been its extension to non-conformal field theories.
It has been appreciated for some time that one way to break conformal
symmetry is to consider, on the string side, branes that wrap vanishing
cycles. In particular, in a beautiful series of works \refs {\kw, \kn,
\gubke , \kt, \ks, \fkt} the gauge theories that arise when one places 
regular D3 branes plus wrapped D5 branes at the simplest conifold 
singularity have been studied, and compared with the corresponding
supergravity duals (see refs{\hko, \revtwo} for reviews).

In particular, in reference \ks\ Klebanov and Strassler (KS),
managed to write a regular IIB supergravity solution corresponding
to the case of having $N$ regular and $M$ fractional D3 branes, whose
dual description involves 4d ${\cal N}=1$ $SU(N+M)\times SU(N)$ gauge 
theories. A key feature of that solution is that the 5-form flux varies in 
the radial
direction of the AdS part of the background. This translates, in the dual
gauge theory, into a running of the gauge couplings. As the couplings 
run, eventually one of them gets strong, and the best description is in 
terms of a similar gauge theory, but with lower rank $N\rightarrow N-M$. 
In \ks , this change in ranks was interpreted as Seiberg
duality \seiberg. In this fashion, the gauge dual of the supergravity
solution is a RG flow, where the effective field theory description
changes an arbitrarily large number of times, as we vary the energy
scale. This phenomenon was dubbed a ``duality cascade'' in \ks .

An important point is that in the presence of wrapped D5-branes, none of 
these $SU(N+M)\times SU(N)$ theories is asymptotically free so, strictly 
speaking, they are not 
well defined quantum field theories at all energy scales. This raises a 
number of interesting questions, e.g. whether the KS gravitational background 
is dual to an $N=\infty$ theory, rather than a finite $N$ gauge theory, or 
whether one can find a UV completion of this RG flow in terms of ordinary 
field theory.

Given the remarkable features of this supergravity solution and its
dual interpretation, it seems natural to try to generalize it to other
examples. For instance, it would be desiderable to understand at a more
general level the relation between the asymptotic (IR in the geometry) 
behavior of  supergravity solutions with varying flux, and the UV properties 
of the field theory dual. Generalizing the KS solution to other 
backgrounds/brane 
configurations does not seem straightforward, but at least some questions
can be addressed without knowing the supergravity solution. Some steps
along that direction were taken in \cfikv\ , where the duality cascade of 
\ks\ was reinterpreted in terms of geometric transitions among vanishing 
2-cycles. This identification allowed for immediate
generalizations, and duality cascades corresponding to transverse D3
branes plus wrapped D5-branes on  ADE fibrations (including ADE conifolds) 
were also discussed in \cfikv\  (see also \tttata).

It has been known since the original work of D-branes on ADE orbifolds 
\refs {\gimon, \dm}, that D-branes at orbifolds have worldvolume theories 
with matter content
dictated by the corresponding ADE Dynkin diagram, and in particular, adding
D-branes transverse to the orbifold corresponds to adding an affine
node to the Dynkin diagram. This is also true for ADE conifolds, so placing
D3-branes transverse to a conifold yields affine ADE gauge theories,
whose Weyl group is infinite. It was pointed out in \cfikv\ that this
infinite order of the affine Weyl group is the reason underlying the 
existence of duality cascades.

In this paper we further explore possible duality cascades, considering
general ${\cal N}=1$ quiver theories, beyond the ADE or affine ADE cases,
from a purely field theoretical point of view. We recast the phenomenon
of duality cascades in terms of the Cartan matrix one can naturally
associate to a quiver gauge theory - which is ${\hat A_1}$ 
in the KS example -, and we argue that the UV behavior of the duality cascade 
depends markedly of the type of Cartan matrix. In this language, Seiberg 
dualities are realized as Weyl reflections \cfikv.

The main novelty of the present work is a discussion of duality cascades
involving quiver gauge theories whose Cartan matrix has exactly one negative 
eigenvalue. In particular we analyze in detail the simplest example: 
${\cal N}=1$ gauge theories with gauge group $SU(N_1)\times 
SU(N_2)$ with $k>2$ chirals in $(N_1,\bar N_2)$ and $k$ chirals in 
$(N_2,\bar N_1)$. As we consider the different flows that arise as we vary 
$N_1$ and $N_2$, some surprises arise, as all the flows seems to converge 
towards the UV, and furthermore, the scales at which we are supposed to 
change the effective description pile up at an accumulation point. This 
behavior had been already observed in some models by M. Strassler, who 
named it a ``duality wall'' \strassler. These properties lead us to speculate 
that for each fixed $k$, a new kind of theory exists (perhaps some 
little string theory), controlling the UV of all these flows. 

The organization of the paper is the following. In section 2 we define
the Cartan matrix of a quiver gauge theory, and relate its Weyl reflections
with Seiberg duality applied to different gauge factors of the quiver theory.
Duality cascades arise as flows connecting theories that are roots of the
same Weyl orbit. In section 3 we review the cases of quiver gauge 
theories with ADE and affine ADE Cartan matrices. These have been extensively 
studied in the literature, as they arise as the worldvolume theory of D-branes
in different singularities. Then we move in section 4 to consider quiver 
gauge theories with a hyperbolic Cartan matrix, and discuss in detail a 
simple example, stressing the qualitative difference in behavior with the KS 
duality cascade. Section 5 contains conclusions and ideas for further work.

\newsec {Weyl reflections and Seiberg dualities.}

We will consider ${\cal N}=1$ gauge theories of the quiver type: the gauge 
group is $G=\Pi_{i=1}^n\; SU(N_i)$, and chiral superfields transform in 
bifundamental representations $(N_i,\bar N_j)$. In the context of string
theory, quiver gauge theories arise as the worldvolume theories of 
D-branes \refs{\dm,\dgm}.  In what follows, we apply to quiver gauge 
theories some 
ideas familiar from the study of Lie algebras, starting with the definition
of the Cartan matrix associated with a quiver gauge theory\foot{ See
\yanghui\ and the appendix of \dfr\ for introductions to quiver theory and 
some of its applications to supersymmetric gauge theories. The Cartan 
matrix associated with a quiver gauge theory has appeared, in a more or less 
explicit way, in a number of works \refs {\kmv, \yanghui, \dfr}. As far as I 
am aware, the associated Weyl reflections were first discussed in the physics
literature in \dfr ; their relation to Seiberg dualities was pointed 
out - in the ADE and affine ADE cases - in \cfikv , although it is implicit
in previous work realizing Seiberg duality through D-branes on Calabi-Yau's 
\refs {\oovafa, \feng, \beas, \tata, \ttata}.}. As we will see, the Cartan 
matrix encodes the matter content of the theory, while theories that differ 
by the ranks of the gauge groups can be thought of as vectors on a root space.
In particular, Weyl reflections change the ranks of the gauge groups
as Seiberg dualities do \cfikv. 

Our starting point is the {\it Cartan matrix} of a quiver gauge
theory with gauge group $G=\Pi_{i=1}^n\; SU(N_i)$, $i=1,\dots,n$, defined as

\eqn\cartan{C_{ij}=2\delta_{ij}-\hbox{\# of chirals in } (N_i,\bar N_j)}

Note that the Cartan matrix thus defined is insensitive to the possible
existence of a superpotential, or to the presence of matter in representations
other than the bifundamental. In the present paper we restrict to ${\cal N}=1$
gauge theories with chiral matter in ${\cal N}=2$ hypermultiplets, namely
for every chiral in $(N_i,\bar N_j)$, there is a chiral in $(\bar N_i,N_j)$.
This means the Cartan matrix just defined is symmetric for these theories.

Once we specify the superpotential, a particular gauge theory is
determined by the ranks of the gauge groups, $(N_1,\dots, N_n)$. We
can think of this n-tuplet of positive integers as a {\it positive root} 
of an algebra associated with the generalized Cartan matrix \cartan . For 
each gauge group $U(N_i)$, the number of colors and flavors, in the limit 
where all the other gauge couplings are weak, is given by

$$N_c(i)=N_i\hskip1cm N_f(i)=\sum _{j\neq i} |C_{ij}|N_j$$

To define the Weyl reflections, first we associate a simple root 
$\vec \alpha _i$, to each node of the diagram. Then, for each simple
root $\vec \alpha _i$ we define an action over an arbitrary root 
$\vec \beta$ of the algebra (in the remaining, there is no summation over $i$)

$$R_i (\vec \beta )=\vec \beta -(\vec \beta \cdot \vec \alpha _i) \vec 
\alpha _i$$

In particular, acting on another simple root,

$$R_i(\vec \alpha _j)=(\delta _{jk}-\delta _{ik}C_{ij})\vec \alpha _k \equiv 
R_{jk}^{(i)}\vec \alpha _k$$

It is easy to see that $R_{(i)}^2=1$, and that $R_{(i)} C=CR_{(i)}^T$. 
Thinking of our gauge theory as a vector $\vec N=N_i\vec \alpha_i$, in the 
"root space" with components $(N_1,\dots ,N_n)$, we are actually more 
interested in taking the passive point of view on the reflection, and
consider its effect on the components of the vector. By requiring 
$\vec N=\vec N'$, we deduce that under the $i$-th Weyl reflection,

\eqn\weyln{N'_j=N_k R^{(i)}_{kj} =(\delta _{jk}-\delta _{ij}C_{ik})N_k}

\noindent
where we have used that $R_{(i)}^2=1$. Note that the only $N$ that changes
under $R_i$ is $N_i$. Also, the length square of 
the vector, $\vec N^2= C_{ij}N_iN_j$, is invariant under Weyl 
reflections, something that we will be using repeatedly in what follows. In 
physical terms, a Weyl reflection sends a theory with gauge group 
$G=\Pi_{i=1}^n \; SU(N_i)$ to another with gauge
group $G=\Pi_{i=1}^n \; SU(N_i')$ and a chiral in $(N_j',\bar N_k')$
for every original chiral in $(N_j,\bar N_k)$. It is revealing to 
write the action of the  $i$-th Weyl reflection on $N_c(i), N_f(i)$,

$$N_c(i) \rightarrow N_f(i)-N_c(i)\hskip1cm N_f(i)\rightarrow N_f(i)$$

We see that the effect on $N_c(i),N_f(i)$ is identical to that
of a Seiberg duality applied on the $i-$th gauge group, which is
also an operation that squares to the identity. For general quiver
theories, it would be wrong to conclude that Weyl reflections 
correspond to Seiberg duality: to start with, the definition of 
Weyl reflections doesn't depend at all on possible superpotentials;
more importantly, in the original Seiberg duality, the dual magnetic 
theory has a massless meson, and a superpotential. Therefore, Weyl
reflections have a chance to correspond to Seiberg duality only for
theories where that massless meson doesn't appear in the dual theory.

It is well known \refs {\ls, \is} that particular choices of the superpotential
give mass to that meson. For a SU(N) theory
with fundamental and antifundamental matter, a quartic superpotential
does the job. Furthermore, it is self-dual, in the sense that the
magnetic theory presents a similar superpotential in the dual quarks.
Luckily, for ${\cal N}=1$ quiver theories, these
quartic superpotentials appear quite naturally. If we start with a 
${\cal N}=2$ quiver theory, it has a superpotential

$$W_{{\cal N}=2}=\sum ' Q_{ij}^I(\phi _j) Q_{ji}^I $$

\noindent
where $\sum '$ means the sum is weighted with the appropiate signs. We can 
deform this theory to a ${\cal N}=1$ theory by adding some superpotentials 
for the adjoint fields

$$W_{{\cal N}=1}=W_{{\cal N}=2}+\sum W_i (\phi_i)$$

If at some energy scale we integrate out the $\phi$'s, we indeed
obtain quartic superpotentials for the $Q$'s. In this paper we 
will be considering examples with such quartic superpotentials.
It would be interesting to explore whether there are other superpotentials
that are self-dual and get rid of the massless meson.

For ${\cal N}=1$ quiver theories, such a quartic superpotential
appears, for instance, when one considers the worldvolume theory
of N D3 branes transverse to a conifold, as discussed by
Klebanov and Witten \kw , and its self-duality plays a pivotal
role in the duality cascade - to be reviewed in section 3.2 - that arises 
when one considers additional wrapped D5 branes \ks . Finally, the 
dualization of a gauge factor in ${\cal N}=1$ quiver theories was 
analyzed in \cfikv , where the Weyl reflected theory was obtained after
duality plus Higgsing, following the analysis presented in the
appendix of \egkrs.

\subsec {Weyl groups and RG flows.}

For a given quiver gauge theory, the set of Weyl reflections \weyln\ 
forms a group - the Weyl group - and in principle, by judiciously acting
with these reflections on a theory, it can be mapped to infinitely many
others in the same Weyl orbit. The question arises of whether these
theories are related dynamically. An example in the literature where 
infinitely many ${\cal N}=1$ quiver gauge theories are related by 
Seiberg duality is the duality cascade of \ks . In \cfikv , this duality 
cascade was recasted in the language of Weyl reflections, and some 
generalizations were presented, although no attempt was made to write down 
the corresponding supergravity solutions. Our aim is to consider further 
generalizations of the original duality cascade; we will argue that many 
features of the flow are neatly encoded in the Cartan matrix \cartan\ of 
the theories entering the cascade, justifying the introduction of this 
formalism.

For general quiver theories, our approach is the following: once the matter
content - i.e., the Cartan matrix - and the superpotential are fixed, the
starting point is a particular gauge theory - i.e., a n-tuplet $(N_1,\dots, 
N_i,\dots, N_n)$ - , an energy scale $\mu$ and some values for the gauge
couplings at that scale $g_i(\mu)$. Requiring that the gauge couplings
are positive fixes whether the scales $\Lambda_i$ are bigger or smaller
than $\mu$. Then we flow towards the IR, by changing $\mu$. Eventually one 
of the gauge couplings will become strong, and 
at that energy scale we perform Seiberg duality - i.e, a Weyl reflection - on 
the particular gauge factor. That yields a new theory $(N_1,\dots ,N_i',
\dots , N_n)$ with only the $N$ of the dualized factor changed. As we
flow to the IR, it makes sense to dualize as long as the electric coupling is
stronger than the magnetic one in the IR. For $SU(N_c)$ with $N_f$ flavors,
this is the case as long as $N_f<2N_c$, i.e. $N_c'<N_c$\seiberg. For the quiver
theories, we also have a superpotential and the other gauge factors, but these
shouldn't change the fact that as we flow to the IR, 
$N_c'<N_c$, i.e. the ranks of the gauge groups keep decreasing.

Although much attention has been devoted to the IR properties of the
KS duality cascade, its UV behavior also raises a number of interesting
questions. Since at each energy scale the effective gauge theory is not 
asymptotically free - as the two beta functions have different sign - it 
is not immediate to find a UV completion of that RG flow, purely in terms 
of ordinary field theory. 

For the KS cascade, the supergravity solution provides the UV behavior;
on the other hand, for some of the examples we will consider later, we 
lack a gravity/string dual description of any flow. Still, we would like
to ask what is the UV behavior of these cascades. Clearly, given an
effective field theory, there is no unique way to flow ``upwards'', towards
the UV. What we propose is to reconstruct a particular flow, by mimicking
the behavior of the KS flow, i.e, let the couplings run towards the UV
and every time a particular coupling gets strong, we perform a Weyl 
reflection. That changes the sign of the corresponding beta function, so 
that coupling now gets weaker towards the UV, allowing us to let the couplings
run upstream until another coupling gets strong, at which point we
repeat the same operation. A point of caution is required: for specific 
examples, dynamically we will see that this prescription calls for applying 
Seiberg duality on a particular gauge factor when the rest of the couplings 
are not weak. We will comment below on to what extend this procedure is 
justified.

Finally, let's mention that some examples of duality cascades for quiver
gauge theories with non-symmetric Cartan matrix were considered in
\cfikv. In the cases 
studied there, the matter content changed at each step of the cascade, so the 
theories appearing in a particular flow don't have a common Cartan matrix. 
Physically, this corresponds to the fact that in these cases some of the 
Seiberg mesons remain massless (as can be anticipated by demanding that the 
dual theory has no anomalies). The UV behavior of such cascades is an
open question left for future work.

\newsec {UV behavior and Cartan matrices.}

The classification of Cartan matrices according to their eigenvalues 
is well established, and we will argue that it dictates the basic
features of duality cascades. The simply-laced positive definite ones 
follow the ADE classification. In string theory, ${\cal N}=1$ or 
${\cal N}=2$ ADE and affine ADE theories can appear as worldvolume theories 
of D-branes \refs {\dm ,\dgm}. From a purely field theoretical 
point of view, it seems worth considering more general quiver gauge theories. 
In particular, we will study quiver gauge theories with a hyperbolic Cartan 
matrix, for which we are not aware of any realization in string theory.

\subsec {ADE theories.}

The classification of positive definite Cartan matrices is familiar
from the study of simple Lie algebras. We pay special attention to
the ones corresponding to simply-laced algebras, namely the ADE matrices.
The Weyl group is finite, so just a finite number of gauge theories
are related by Weyl reflections in this case.

For ${\cal N}=2$, these theories can be obtained in type II string
theory by placing fractional branes on a $\IC^2/\Gamma$ orbifold, if 
we allow only fractional but no regular branes. Recent work on finding
the supergravity duals for these theories is nicely summarized in
\italians . We can break these ${\cal N}=2$ theories to ${\cal N}=1$ by 
adding some superpotential for the adjoint fields. In the case of quadratic 
superpotentials $W(\phi _i)\sim \phi_i^2$, these
${\cal N}=1$ theories appear as worldvolume theories of fractional branes
in the $ADE$ conifolds considered by \gns . More general superpotential
deformations appear by considering branes in other ADE fibrations \cfikv.

\subsec {Affine ADE theories and duality cascades.}

Consider a quiver theory whose Cartan matrix is an affine ADE matrix, i.e. 
one of the eigenvalues is zero and the rest are positive. If we denote by 
$d_i$ the Dynkin indices 
of the affine Dynkin diagram, the eigenvector with zero eigenvalue is $(Nd_0, 
Nd_1,\dots , Nd_r)$. We have to differenciate between gauge theories 
corresponding to the zero eigenvector $(Nd_i)$ and the rest. 

i) In all the cases considered so far in the literature, the theories with 
$G =\Pi_{i=1}^n U(Nd_i)$ flow to a non-trivial IR fixed point. In the context 
of string theory, these theories appear as the worldvolume of $N$ regular 
branes placed transverse to a $\IC^2/\Gamma$ singularity (for ${\cal N}=2$) 
or an ADE fibration (for ${\cal N}=1$).

For ${\cal N}=2$ theories, the $\beta $ functions are exactly zero.
These theories are discussed in \refs {\kmv, \lnv}. For ${\cal N}=1$ theories,
the $\hat A_1$ case was considered by Klebanov and Witten \kw, who studied 
the worldvolume theory of N D3 branes transverse to the $A_1$ conifold. 
The gauge theory they obtained is $U(N)\times U(N)$ with 2 chirals in 
$(N,\bar N)$ and 2 chirals in $(\bar N,N)$, so indeed the Cartan matrix is 
$\hat A_1$ and $(N,N)$ is the vector with zero eigenvalue.

The generalization to $\widehat {ADE}$ theories with quadratic
superpotentials $W(\phi _i)=m_i\phi^2$ was studied by \refs{\gns,\lopez}. 
The theories that arise in the discussion of $\IZ_k$ orbifolds of the 
conifold \uranga\ turn out to have affine $A_{2k-1}$ Cartan matrix. In 
\cfikv\ it was argued that the ${\cal N}=1$ theories
with gauge group $G=\Pi_{i=1}^n U(Nd_i)$ and superpotentials $W(\phi_i)=
\phi ^{k+1}$ also present IR fixed points, for arbitrary $k$.

ii) The case when the gauge group is not the zero eigenvector is
realized in string theory by placing both fractional and regular
branes at the orbifold or ADE conifold (or ADE fibration). In this case 
we have an arbitrarily large number of theories
related by a RG flow. To characterize the possible Weyl orbits, recall that 
$\vec N^2$ is invariant under the flow; in the affine case we have 
$\vec N^2 \geq 0$. Embedding the 'root space' $\IN^n$ in $\IR^n$ and allowing 
momentarily the $N's$ to be real, the equation $\vec N^2= s^2$ describes 
ellipsoids times $\IR$. The degenerate case $s^2=0$ 
corresponds to a straight line. Weyl orbits consist of points with integer 
coefficients in these hypersurfaces.

In the $\hat A_1$ case, the supergravity solutions have been written down: 
for ${\cal N}=2$, a (singular) supergravity 
solution corresponding to N D3 branes and M wrapped D5 branes on 
$\IC^2/\IZ_2$ appeared in \polchi. The physics of that ${\cal N}=2$ RG 
flow is not expected
to be a duality cascade, but rather a series of Higgsings \ofer. For 
${\cal N}=1$, Klebanov and Strassler \ks\ constructed a 
regular IIB supergravity solution, obtained by considering $N_1$ D3 branes 
transverse to the conifold singularity, plus $M$ D5 branes wrapping a
$S^2$ cycle. Defining $N_2=N_1+M$, these theories are of the form 
$SU(N_1)\times SU(N_2)$ with 2 chirals in $(N_1,\bar N_2)$ and 2 chirals in 
$(\bar N_1, N_2)$ - so indeed its Cartan matrix is $\hat A_1$. 

Let's illustrate the formalism introduced with the KS example. In this case 
$\vec N^2=2(N_1^2+N_2^2-2N_1N_2)=
2M^2$, which tells us that $|M|$ is constant along the flow. The orbits
reduce to $S^0\times \IR$, i.e. two straight lines in the $N_1\times N_2$
plane, given by $N_2=N_1\pm M$. As we flow towards the IR, we reach a theory
for which, if we tried to apply yet another Weyl reflection we would run 
into the region with negative $N$'s. This is the IR endpoint of the cascade,
and for fixed $M$ we have $M$ possible such fixed points, with theories
$SU(M+p)\times SU(p)$, $p=1,\dots,M$. The physics depends on the ratio $p/M$
\ks : for $p\ll M$ KS argue that there is confinement for some range of energy 
scales.

\ifig\flowaff{Affine flow of the Klebanov-Strassler duality cascade.
Different flows can have the same value of $M=N_2-N_1$; their IR endpoints
are represented by the dots.}
{\epsfxsize2.0in\epsfbox{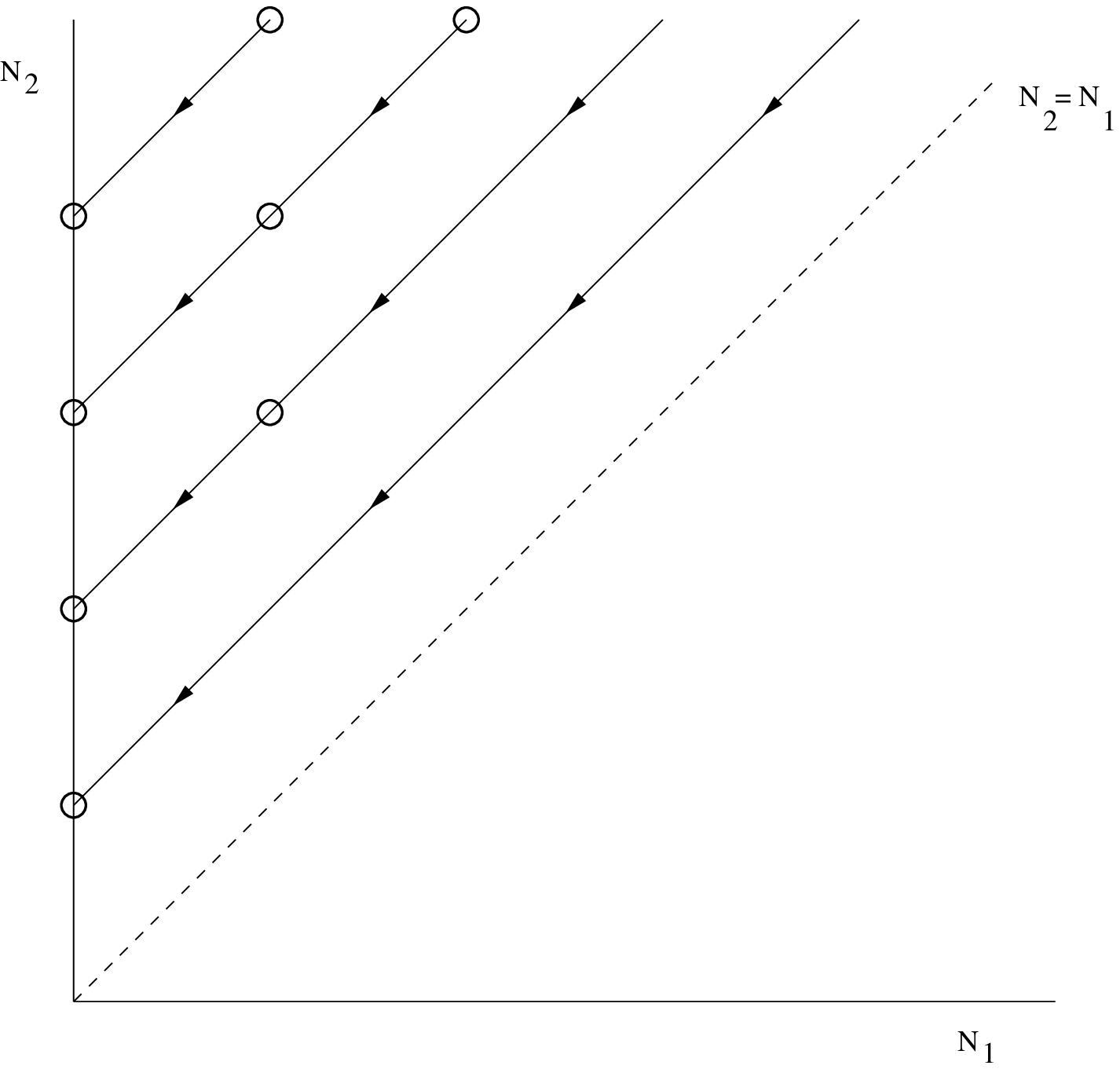}}

In \flowaff\ we plot these flows in the $N_1-N_2$ plane. Using the 
$\IZ_2$ symmetry between the two gauge factors, we represent theories 
choosing always $N_2>N_1$.

Another feature of the KS duality cascade is that towards the UV
the $N's$ grow linearly with the number of steps in the cascade,
namely, after $n$ steps, $SU(N)\times SU(N+M)$ has turned into
$SU(N+nM)\times SU(N+(n+1)M)$. This ``linear growth'' of the N's is general 
for all affine ADE quiver theories, since for affine $ADE$ - except 
$\hat A_1$ - there is at most one hypermultiplet transforming under any 
two gauge factors, so after $n$ steps, if we rewrite the new N's in term 
of the original ones, each coefficient is going to be at most $n$.
Finally, for future comparison, let us remember that in the KS cascade,
the couplings behave periodically. 

\ifig\flowaff{Periodic behavior of the two gauge couplings in
the Klebanov-Strassler duality cascade.}
{\epsfxsize2.0in\epsfbox{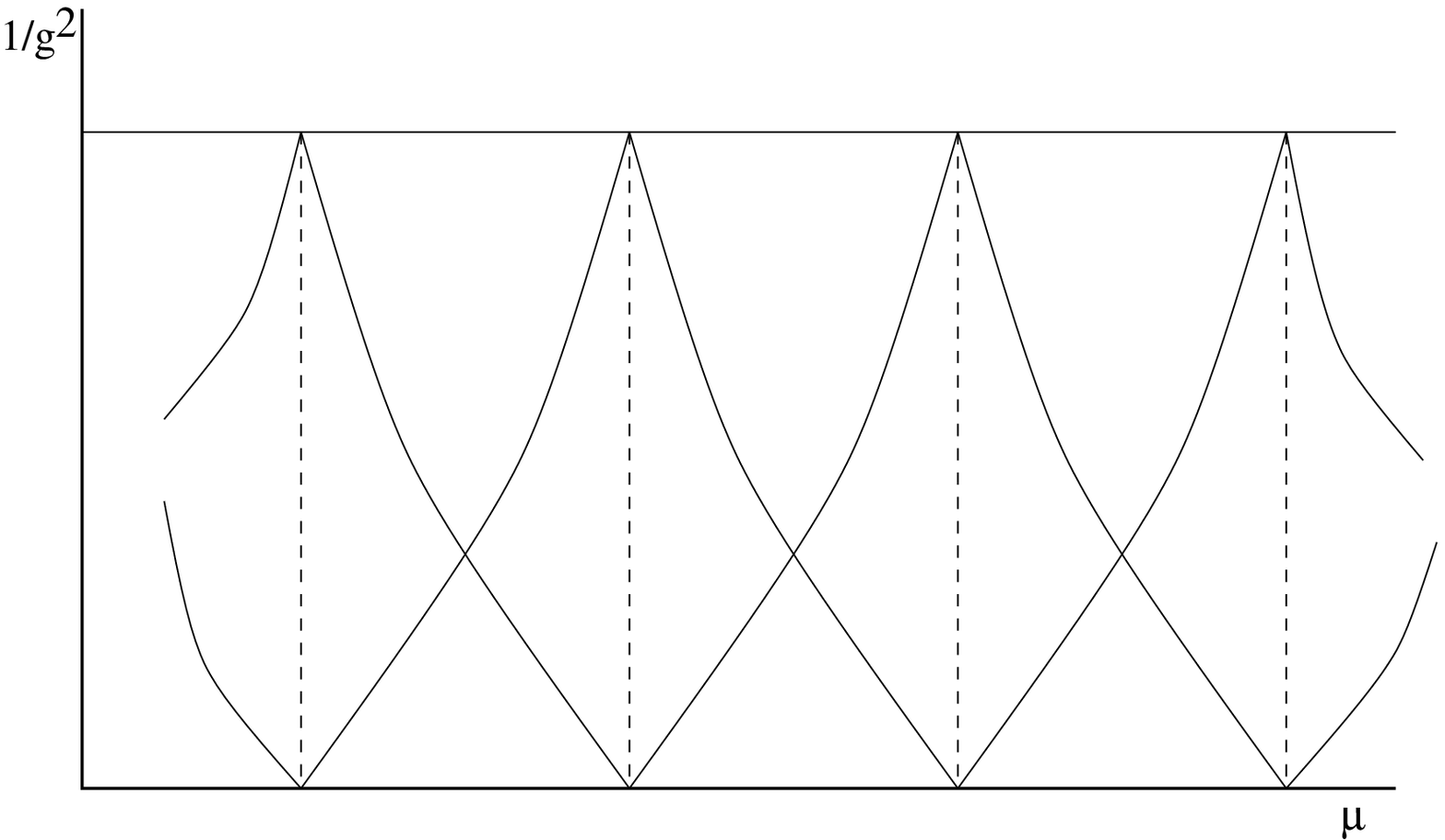}}

\newsec {Hyperbolic theories and duality walls.}

Once we leave the realm of ADE or affine ADE Cartan matrices, there
is no full classification of the possible Cartan matrices. For the
sake of concreteness, we will restrict in the following to gauge theories
with a hyperbolic Cartan matrix, i.e., matrices with a single negative
eigenvalue and all the other positive. These bilinear forms are familiar
from General Relativity, and in classifying such theories we can borrow 
some of the language of GR. Contrary to the previous cases, we are not aware 
of any realization of these theories as world-volume theories for D-branes, 
as this would amount to engineer configurations with an arbitrary number of 
hypermultiplets in a given bifundamental representation. After a few 
generalities, we will study in some detail the simplest example.

A first novelty, compared with the affine theories, is that the equation 
$\vec N^2=s^2$ can now have positive, zero or negative $s^2$ (we could 
call them space-like, light-like and time-like theories).
Again, allowing momentarily the $N$'s to be real, these equations correspond 
to hyperboloids in $\IR^n$. The degenerate case $\vec N^2=0$ corresponds to a 
cone going through the origin (the light-cone). 

Applying, as in \kmv , the Perron-Frobenius theorem to the adjacency matrix 
$A=2I-C$, we learn that the eigenvector with smallest eigenvalue of the 
Cartan matrix can always be chosen to have all entries positive. For a 
hyperbolic matrix, the smallest eigenvalue is the negative one, so we deduce 
that the eigenvector with the negative
eigenvalue (the one giving the ``time direction'') can be chosen to have
all entries positive, and it corresponds to a sensible gauge theory.
Also, note that the axis $(0,\dots,N_i,\dots ,0)$ belong to the 
$\vec N^2>0$ region. 

We want to argue that the typical IR behavior of flows with $\vec N^2>0$ and
$\vec N^2<0$ is different. Before proceeding with the argument, it will be
useful to introduce a couple of definitions. First, for a given vector 
$\vec N$, we define the vector $\Delta \vec N$ with components
$\Delta N_i=\left (R^{(i)}(\vec N)\right )_i-N_i=N_i'-N_i$, i.e. its $i$-th
component encodes how much the vector $\vec N$ would change, were we to
perform the $i$-th Weyl reflection on it. A useful observation is that,
by \weyln , we can write $\vec N^2=-N_i\cdot \Delta N_i$, where $\cdot$ is
the ordinary inner product in $\IR^n$ (no Cartan matrix involved).

We are now ready to discuss the difference between $\vec N^2>0$ and
$\vec N^2<0$ flows. Given a Weyl orbit, there is a minimum distance - in
the canonical Euclidean metric of $\IR^n$ - from the points representing 
theories in that orbit to the origin of $\IR^n$. If a theory corresponding 
to such minimum has all N's positive, then for any Weyl reflection acting 
on it, $N'_i>N_i$, i.e. $\Delta \vec N$ has all components positive, and 
from the observation in the previous paragraph, we deduce that it must 
correspond to a $\vec N^2<0$ flow. We then conclude that if all the $N's$ 
are  positive at a minimum, it must be an orbit with $\vec N^2<0$, or 
equivalently, all the $\vec N^2>0$ orbits have $N's$ of mixed signs at the 
minimum.

Physically, we interpret this difference as follows. Recall that as we
flow towards the IR, the $N'$s decrease. On the $\vec N^2>0$ region, 
since the minimum of the orbit has $N's$ of mixed signs, as we 
flow towards the IR, the flow eventually will enter the regions where some 
of the ranks  of the gauge groups would be negative, so they make no sense 
anymore, exactly as in the KS solution. As in the KS cascade, the IR endpoint
for $\vec N^2>0$ flows is reached when the strongly coupled gauge group has
$N_f(i)<N_c(i)$. Again, as in KS, the physics of the IR endpoint will depend 
on the rest of the $N$'s (what was called $p$ in the previous section), but
for endpoints very near the axis of $\IZ^n$ we expect again confinement
for some range of energy scales (for theories on the axis, pure $SU(N)$,
we expect confinement at all scales).

For the theories in the $\vec N^2<0$ region, there will be some flows with the
behavior just described, but the typical flow in this region will have a 
different IR behavior. Now the IR endpoint is characterized by $N_c(i)'>
N_c(i)$ for all $i$. In the example we will consider next, we find that
IR endpoint can be a non-trivial CFT, or IR free theories.

\subsec {An example.} 

Let's illustrate these points with a simple example. Consider
${\cal N}=1$ $d=4$ gauge theories with gauge group $SU(N_1)\times
SU(N_2)$ with $k>2$ chiral multiplets $A_i$ in $(N_1,\bar N_2)$, $k$ chirals 
$B_i$ in $(N_2,\bar N_1)$ and quartic superpotential

\eqn\superpot{W=\lambda \epsilon ^{i_1i_2i_3\dots i_k}\epsilon _{j_1j_2i_3
\dots i_k}\hbox {Tr }A_{i_1}B^{j_1}A_{i_2}B^{j_2}}

\noindent
with a $U(k)$ global continuous symmetry. The Cartan matrix is

$$C=\left(\matrix{2& -k\cr -k& 2}\right)$$

\noindent
with eigenvectors $(1,-1)$ and $(1,1)$ (the ``time direction'').
The Weyl reflections, acting on the right on the row vector 
$(N_1\; N_2)$, are

$$R_1=\left(\matrix{-1& 0\cr k& 1}\right)\hskip1cm 
  R_2=\left(\matrix{1&k\cr 0&-1}\right)$$





The $\vec N^2=0$ cone can be described by defining $\lambda = N_2/N_1$, 
and rewriting the previous equation as $\lambda ^2-k\lambda +1 =0$, which 
has the solutions

$$\lambda _{\pm}={k\pm \sqrt {k^2-4}\over 2}$$

Note that $\lambda_- =1/\lambda_+$, which corresponds to the exchange
of $N_1$ and $N_2$. The straight lines $N_2=\lambda _\pm N_1$ give the
$\vec N^2=0$ cone in $\IR^2$, and the generic hyperbola $\vec N^2=2(N_1^2+
N_2^2-kN_1N_2)=s^2$ asymptotes towards this cone. In this example, all
the theories with $\vec N^2<0$ have both $N$'s positive,
and correspond to sensible gauge theories. We eliminate the $\IZ _2$ 
symmetry by always considering theories with $N_1<N_2$.

\ifig\flowhyper{Hyperbolic flow for theories $SU(N_1)\times SU(N_2)$
with $k>2$ hypermultiplets. After modding out by the $\IZ_2$ symmetry,
the $N_2=\lambda _+ N_1$ represents the ``light-cone'' separating the
$\vec N^2 >0$ and the $\vec N^2 < 0$ regions.}
{\epsfxsize2.0in\epsfbox{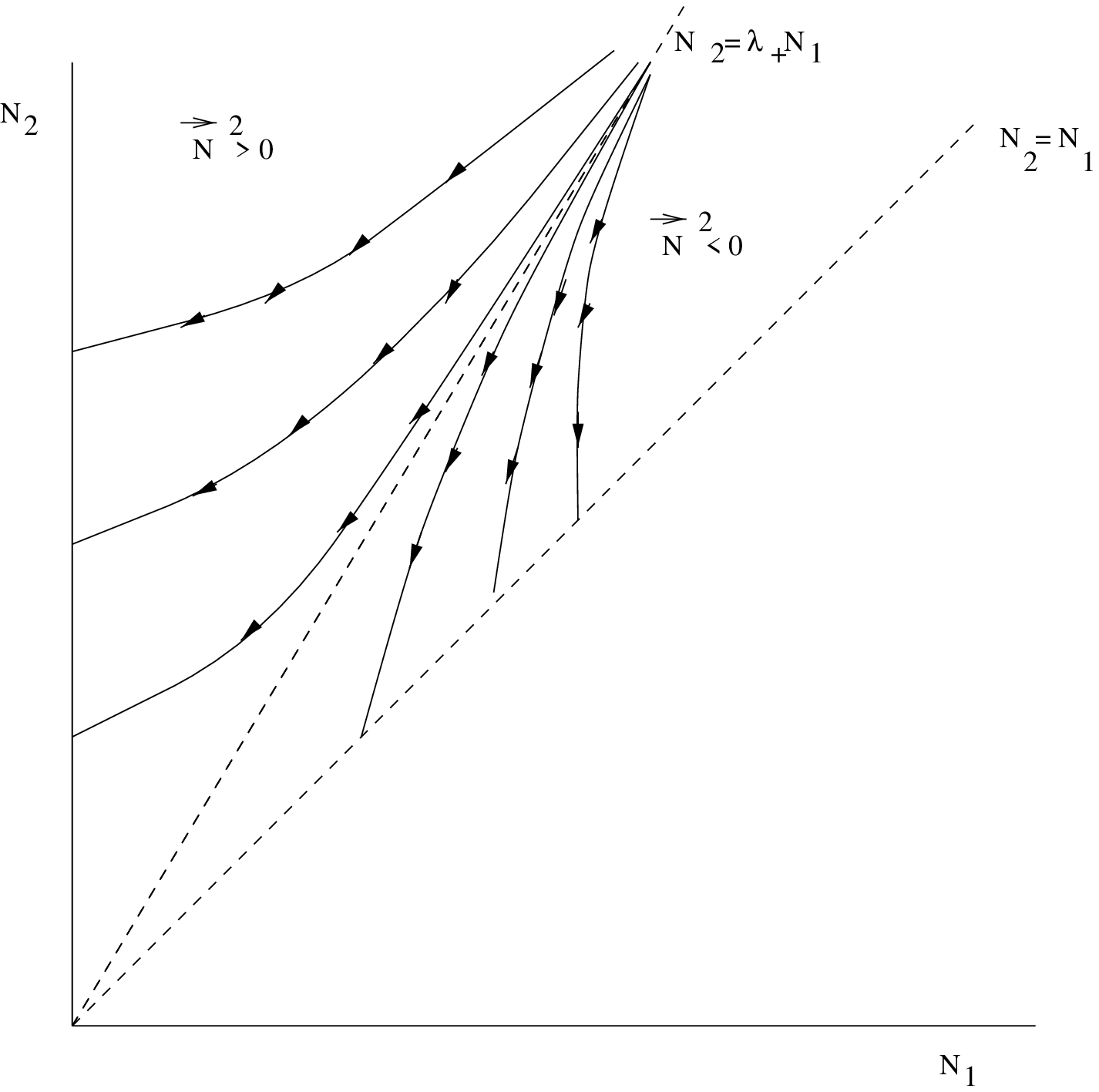}}

Since primitive Weyl reflections square to the identity, if we want to
have a RG cascade relating an infinite number of theories, we should
ensure that we are in a regime where the couplings run in such a way
as to require alternate dualizations, as was the case in the KS
duality cascade. Namely, we would like to determine under which conditions 
the beta functions for the two groups have different signs. As in \ks\ , we 
will use the beta functions for inverse square couplings, 
with holomorphic gauge kinetic term and canonical matter kinetic term,
\foot{In footnote 3 of \hko\ it is pointed out that in the KS cascade, this
choice is the natural one as dictated by supergravity.}

\eqn\betafun{\beta _1 \left ({8\pi^2\over g_1^2}\right )=3N_1-kN_2(1-\gamma) 
\hskip1cm \beta_2\left({8\pi^2\over g_2^2}\right)=3N_2-kN_1(1-\gamma)}

with $\gamma$ the anomalous dimension of the $\hbox {Tr }(A_iB_j)$ 
operators, common to all of them due to flavor symmetry. Here a 
difference appears, compared with the KS solution reviewed in the 
previous section. In KS it was crucial to take into account the effect of the
anomalous dimension $\gamma$ of $\hbox {Tr }(A_iB_j)$ on the
beta functions of the two gauge couplings, to obtain beta functions with
different sign. In the present theories, before taking into account the 
presence of anomalous dimensions, it is straightforward to find a range of 
$N_{1,2}$ such that the beta functions for the gauge couplings have different 
signs. Indeed, as we will show shortly, all the effective theories appearing
at different energy scales of the cascade  - except the last one in the IR - 
have two gauge factors that are not in their respective conformal windows, 
but rather one is magnetically free in the IR, and the other electrically 
free. The concern is then that the anomalous dimensions don't spoil this 
property.

In general, we can't compute the anomalous dimensions that appear
in the beta functions for the different gauge couplings. For the theories 
appearing in the KS solution, the strategy was the following: first 
consider the theories that flow to a nontrivial CFT in the IR (M=0, no 
fractional branes). For these theories, using the arguments of \ls ,
it is possible to determine the anomalous dimension at the conformal point, 
$\gamma(Q\bar Q)=-1/2$ \kw. Then, for the theories $SU(N+M)\times SU(N)$, 
it was argued that the anomalous dimensions are given by a power series in 
$(M/N)^2$, $\gamma (Q\bar Q)= -1/2+{\cal O}(M^2/N^2)$ \foot{In subsequent 
work \fkt , it was argued that for $M\ll N$ and large 't Hooft 
couplings, the leading correction to the anomalous dimension takes the
form $\left ({M\over N}\right )^4 \left ({1\over g_1^2N}+
{1\over g_2^2N}\right )^{1\over 2}.$}.

Although we are not able to compute the anomalous dimensions 
$\gamma (g_1,g_2,\lambda)$, for the arguments we will presenting, it 
suffices to bound their value. To bound 
the value of the anomalous dimensions from below, we argue as follows\foot{I
would like to thank M. Strassler for helpful discussions on this point.}.
First recall that for ${\cal N}=1$ $SU(N_c)$ with $N_f$ flavors in the 
conformal window, the anomalous dimension of the meson at the IR fixed point 
is given by $\gamma (g_*)=1-3N_c/N_f$ \exact , so in this window, 
$-1\leq \gamma (g_*) \leq 0$. At the lower end of the conformal window, 
$N_f=3 N_c/2$, the dimension of the meson $D(\bar Q Q)$, reaches its minimum 
value allowed by unitarity $D(\bar QQ)=1$, so it becomes free. It is then 
expected that for $N_f\leq 3N_c/2$, $\gamma_{IR}(\bar QQ)=-1$. In the case
we are studying, the gauge factor that is getting strongly coupled towards the
IR is always magnetically free, and although in addition we now have another 
gauge factor and the quartic superpotential \superpot, we are going to assume
that in the strong coupling regime $\gamma \geq -1$.

To bound the anomalous dimensions from above, note that if the quartic
superpotential \superpot becomes marginal, then $\gamma (\bar QQ)=-1/2$
in the IR. If $\gamma (\bar QQ)>-1/2$, the superpotential is an irrelevant
operator, and in the IR we would have the original Seiberg duality, with
the unwanted massless meson. The consistency of the analysis thus requires
$\gamma (\bar QQ)\leq -1/2$. We expect the superpotential to
become marginal for the theories with group $SU(N)\times SU(kN/2)$, since 
for these theories, the strongly coupled gauge factor, $SU(kN/2)$, has 
$N_f=2N_c$, and if we apply Seiberg duality, the full theory goes back to 
itself, which we interpret as evidence of a non-trivial CFT in the IR. Then, 
in the IR the quartic superpotential
becomes a marginal operator and $\gamma (\bar QQ)=-1/2$. Since for 
$N_2/N_1>k/2$ the theory is ``more strongly coupled''
in the IR, by analogy with the $SU(N_c)$ case reviewed above, we expect
the anomalous dimensions in the strong coupling to be smaller than $-1/2$.
Therefore, for theories with $N_2/N_1>k/2$, we are going to assume that
at the scale we perform the duality, $\gamma \leq -1/2$. Luckily, as we will
shortly show, all the flow (except the very last duality transformation 
in the IR) takes place among theories with $N_2/N_1>k/2$, so this bound
suffices for our purposes. It is worth pointing out that most of the arguments
we are going to present would carry through with a more relaxed version of 
the bounds, $-1\leq \gamma \leq 0$, so we are confident that the results
we present are robust.

Our starting point is then a theory with $(N_1,N_2)$ with, say, $N_2>N_1$.
For reasons that will become clear below, we further require $N_2/N_1>k-1$;
this condition is enough to guarantee that $\beta_2>0>\beta_1$. Since
$k-1<\lambda _+$, that theory can be either in the $\vec N^2>0$ or the
$\vec N^2<0$ region. Also, $\beta _2>0>\beta_1$ and the positivity of the 
couplings require that if we are at an energy scale $\mu$, the respective 
scales satisfy $\Lambda _2<\mu<\Lambda_1$. Having established the setup, we 
are going to study the behavior of such theory as we flow towards the IR. 
Later we will reconstruct a flow backwards, towards the UV.

As we flow to the IR, and let the couplings run using \betafun , the gauge 
coupling of the larger group, $SU(N_2)$, gets stronger, while the gauge 
coupling of the smaller gauge group gets weaker. Eventually, we reach a 
scale where $1/g_2^2=0$ and we dualize the larger group (i.e., we apply
$R_2$), so $N_1'=N_1$ and $N_2'=kN_1-N_2$. The new scales are

$$\Lambda _1'=\Lambda _2\left ({\Lambda _1\over \Lambda _2}
\right )^{\beta _1\over \beta_1'}\hskip1cm \Lambda_2'=\Lambda_2$$

Concerning the new $\beta$ functions for the inverse couplings, it is easy 
to argue that from $\gamma '\leq -1/2$, it follows that $\beta'_2<0$, so 
the dualized gauge group now runs towards weak coupling. On the other hand, 
the behavior of the spectator gauge group needs a separate discussion.

For theories with $\vec N^2>0$, just
assuming $\gamma'\geq -1$ suffices to show that $\beta_1'>0$, so the
roles of the two gauge groups get exchanged. This scenario is qualitatively
similar to the KS cascade. As we keep dualizing the bigger gauge factor at 
each step, eventually we reach a point where a further transformation would 
yield $N_2'<0$, so we have reached the IR endpoint. This happens for 
$N_2>kN_1$, and as in KS, for theories near the axis, we expect that the 
strongly coupled gauge factor presents confinement at some energy scales.

For the $\vec N^2<0$ theories, as we apply 
$R_2$ three things can happen, depending on the ratio $N_2/N_1$: 

i) $N_2'<N_1$, which happens for $\lambda_+>N_2/N_1>k-1$, 

ii) $N_2>N_2'>N_1$, which happens for $k-1>N_2/N_1
>k/2$, 

iii) $N_2'>N_2$, which happens for $k/2>N_2/N_1>1$.

In the first case, we go to a theory with smaller ratio (even after the
$\IZ_2$ reflection), so we move away from the $\vec N^2=0$ cone. Again,
assuming $\gamma'\geq -1$, we can conclude that $\beta_1'>0$ and new
couplings still run in opposite directions. However, we are no longer
guaranteed that the new theory still is in region $i)$. Indeed, one can check
analitycally - and it is obvious from \flowhyper\ - , that the slopes
get smaller at a faster rate as the $N$'s decrease, so they eventually
reach a theory in region ii). 

Once that happens, the next point is already in region iii). Note that here 
$N_{1,2}'>N_{1,2}$. This means that both reflections would mean flowing 
upwards (in different trajectories, with the same $s^2$). This is the
IR endpoint of the flow. As we argued before, in this range we expect the 
superpotential to become an irrelevant operator, so we can forget about it
in the analysis of the IR endpoint. We have then two different scenarios: for 
$k/2\geq N_2/N_1\geq k/3$, the theory flows to an interacting CFT, with the
spectator group playing the role of a weakly gauged flavor symmetry; for 
$k/3>N_2/N_1>1$ both groups are IR electrically free.

The upshot of this discussion is that for flows with $\vec N^2<0$, all
the effective theories in the cascade, except the last two, satisfy $\lambda_+>
N_2/N_1> k-1$. Then, near the end of the flow there is a theory with 
$k-1>N_2/N_1>k/2$ and finally a theory with slope $k/2 >N_2/N_1>1$, which 
describes the IR endpoint of the flow.

Having described the flow to the IR, now we would like to study the UV 
behavior of these flows. What is meant by this is that we let the couplings
run towards the UV using \betafun . Eventually one of the couplings will
get strong, signalling that the corresponding gauge factor is not 
asymptotically free, and we propose to produce a new effective theory, valid
at higher energy scales, by applying Seiberg duality on that node.

Again, start with a theory with $N_2/N_1 > k-1$, and $\beta _2 > 
0 > \beta _1$. Now, letting the couplings run upstream
according to \betafun, eventually $1/g_1^2=0$, and we apply Seiberg duality 
to the first gauge factor (i.e., $R_1$) to obtain the new gauge groups and 
beta functions to keep flowing towards the UV. It is easy to argue that as we 
flow back to 
the UV, the gauge couplings take turns in satisfying $ 1/g_i^2 =0$, so we 
have to apply $R_1$ and $R_2$ alternatively. Indeed, after applying $R_1$, 
the new gauge theory has $N_1'=kN_2-N_1$, $N_2'=N_2$ and since we assume 
$-1\leq \gamma' $, it follows that after the reflection we have 
$\beta _1'>0>\beta_2'$. 

The ranks of the gauge groups form then a series $N_1,N_2, \dots N_i, \dots$, 
so at each step of the cascade we have gauge group $U(N_n)\times U(N_{n+1})$. 
The generic term of the series is

$$N_{n+2}={\lambda _+^{n+1}-\lambda _-^{n+1}\over \sqrt {k^2-4}}N_2-
{\lambda _+^n-\lambda _-^n\over \sqrt {k^2-4}}N_1$$

After many steps up in the cascade, this can be approximated by 
$N_{n+2}\sim c_1\lambda _+^n$, so the ranks of the gauge
groups grow {\it exponentially} with $n$, the number of steps in the
cascade. This is to be contrasted with the linear growth of the KS
cascade and the affine cascades in general. As a check, from this 
expression it follows that

$$N_{n+2}-\lambda _+ N_{n+1} \rightarrow 0 
\hskip0.5cm \hbox {as } n\rightarrow \infty $$

\noindent
independently of $N_1$ and $N_2$, so all the flows asymptote in the UV to 
the straight line $N_2=\lambda _+ N_1$. 

Now we would like to study the behavior of the scales at which one applies the 
Weyl reflection. On this score we will encounter a surprise. We apply $R_1$ 
at the energy scale when ${1\over g_1^2} =0$, i.e. $\mu =\Lambda _1$. 
Matching the scales, we obtain

$$\Lambda _1'=\Lambda _1\hskip.5cm \Lambda _2'=\Lambda_1
\left ({\Lambda _2\over \Lambda _1}\right )^{\beta _2\over \beta_2'}$$

Running the couplings with the $\beta $ functions \betafun , now towards
the UV eventually we reach ${1\over g_2^2}=0$, and
then apply $R_2$. The net effect of these two reflections is

$${\Lambda _1''\over \Lambda _2''} =\left ({\Lambda _1\over \Lambda _2}
\right )^{{\beta _1'\over \beta _1''}{\beta _2\over \beta _2'}}$$

\ifig\flowaff{Accumulation of scales for the hyperbolic flow. Towards the
UV, the duality cascade can't be extended indefinitely, reaching a 
``duality wall'' at a finite scale $\Lambda _c$.}
{\epsfxsize2.0in\epsfbox{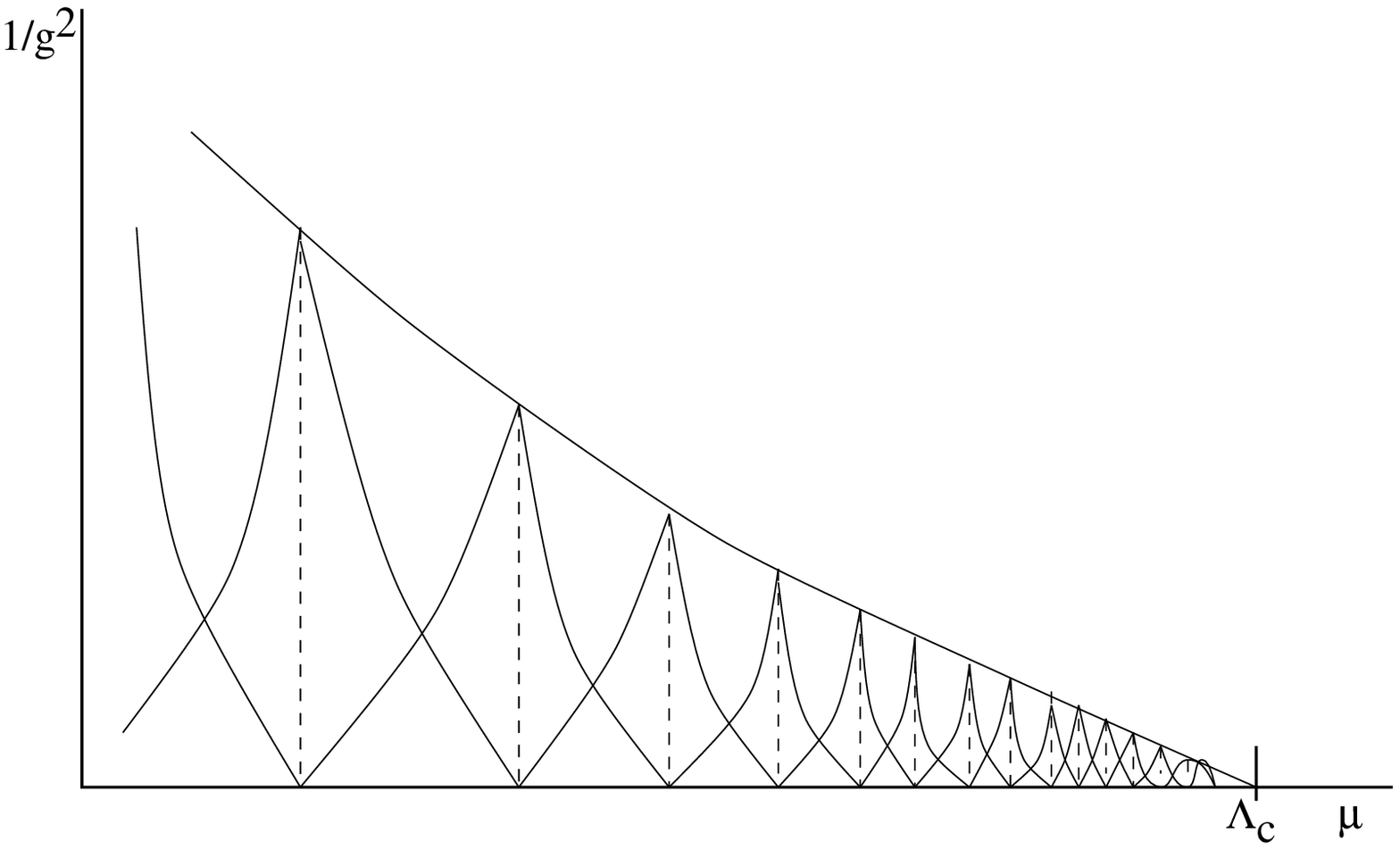}}

As mentioned, now we are guaranteed that the $\beta $ functions run with 
opposite signs, so this procedure can be iterated indefinitely.
There are a couple of things to notice: first, as we flow upwards, towards
the UV, the couplings are getting stronger, contrary to the KS solution,
where the couplings were periodic. Second, the scales at which we perform
the duality also get closer among themselves. Indeed, we can bound the 
exponent that appears in the transformation

$$\left | {\beta _s\over \beta _s'}\right | < 
{2\lambda_+-k\over (k^2-2)\lambda _+ -k} \equiv r < 1$$

where $\beta_s$ denotes the beta function of the spectator gauge group
at each Seiberg duality, and we have used that $N_2/N_1\rightarrow 
\lambda_+$. It follows that after $n$ steps

$$\Lambda _n < \Lambda _1 \left ({\Lambda _1\over \Lambda _2}
\right )^{r-r^{n+1}\over 1-r}$$

and as we keep dualizing there is a finite scale that we can't go beyond of

$$\Lambda _c \sim \Lambda _1 \left ({\Lambda _1\over \Lambda _2}
\right)^{r\over 1-r}$$

The scales pile up towards an accumulation point, defined by a single 
scale $\Lambda_c$. This is quite different from the UV behavior of
the KS cascade, which in principle could be extended to arbitrarily
high energies. This behavior - mentioned at the end of \ks\ - was 
observed by Strassler in some models, and he dubbed it as a ``duality 
wall''\strassler. We will discuss the possible meaning of that limiting
scale in the next section.

Since both gauge couplings are getting strong, pretending that one of the gauge
groups is a spectator and applying Seiberg duality to the other is less 
and less well motivated as we flow back to the UV. Naively we could try to 
appeal to holomorphicity in $\Lambda _1/\Lambda_2$: since Seiberg duality 
applies when $\Lambda _1/\Lambda _2 \ll 1$, we might expect that still holds 
as we let $\Lambda _1/\Lambda_2 \sim 1$. The KS 
supergravity solution offers an example where such argument holds. 
On the field theory side, one of the two gauge couplings ought to be very
weak in order that we can confidently apply Seiberg duality to the other
factor. On the other hand, the supergravity solution is reliable when 
$g_sM\gg 1$. Since $1/g_1^2+1/g_2^2=1/g_s$, in the supergravity description 
of the cascade, the gauge couplings can't be arbitrarily weak, but the
supergravity solution indicates that in that regime, the ``continuation''
of Seiberg duality still holds.

Additional examples of the validity of such continuation are the duality
cascades, both chiral and not chiral, studied in \cfikv. There, Seiberg 
duality is realized in the D-brane setup as monodromy in the Teichm\"uller 
space of the string theory background. Again, in those examples, from the 
field theory point of view, naive application of Seiberg duality is not fully 
justified once all couplings get strong, but thanks to the string theory 
realization, one is confident that indeed the theories form part of a duality 
cascade. 

\subsec {A UV completion?\foot {I would like to thank O. Aharony and
M. Berkooz for helpful comments on a preliminary version of the
following discussion.}}

Given the observations that different flows converge in the UV, and that
the energy scales pile up at a finite energy scale $\Lambda _c$, it is 
tempting to speculate that for every $k$ there exist a new theory, which 
serves as a
UV completion of the flows studied in this section. Borrowing language
from CFT, this theory would be a ``fixed point'' (not in the ordinary
sense, since it has a mass scale), that could be perturbed by a relevant 
operator -with coefficient depending on $s^2$ - triggering the 
different RG flows. For these flows, at different scales the effective 
descriptions would be given by the $SU(N_1)\times SU(N_2)$ theories.

Throughout this section, the examples we will have in mind are those of
4+1 and 5+1 SYM with sixteen supercharges \itzhaki. For each of these two
effective theories, a possible UV completion is provided in the context of 
D-branes in string theory, although ultimately these completions don't 
involve gravitational physics. These UV completions are respectively given
by a 6-dimensional CFT - the (2,0) theory - compactified on a circle, and a 
6-dimensional non-local non-gravitational field theory, the (1,1) little 
string theory. Of course, 
on top of these scenarios - a CFT or some kind of little string theory - we 
can't rule out the possibility that simply there is no UV completion of 
the previous RG cascade. We are going to assume that such UV completion 
exists, and try to guess some of their properties.

First, all the effective theories appearing in a given flow have a continuous
$U(k)$ global symmetry, times a discrete symmetry that, for fixed $k$,
depends on the particular flow (e.g. for KS, the discrete global symmetry 
is $\IZ_{2(N_1-N_2)}$). The natural guess would be that the theory providing
the UV completion has a global $U(1)$ which is broken to discrete subgroups 
when we turn the relevant operator that triggers the flow. Thus, the global 
symmetry group ought to be $U(k)\times U(1)$. Also, all the effective
theories that enter the cascade have 4 supercharges. If the UV completion
is a higher dimensional $(d=5,6)$ CFT, it has at least 8 supercharges, so
the relevant operator that triggers the cascade ought to break some
supersymmetry. 

What would be the asymptotic density of states of these hypothetical
theories? It is straightforward to derive the growth of the ranks 
of the gauge groups as we approach $\Lambda _c$; near $\Lambda _c$
we have $SU(N)\times SU(\lambda_+N)$ with

$$N\sim \left ({\Lambda _c\over \Lambda _c -\Lambda}
\right )^{ln ~\lambda _+\over |ln ~r |}$$

The hard question is, of course, to estimate the number of degrees of
freedom of the quiver theories when both couplings are strong.
We could wonder whether the number of degrees of freedom remains constant 
as we approach $\Lambda _c$, but this seems to be against the 
spirit of dualizing to improve the UV behavior of a not asymptotically
free theory, so the possibility that the number of degrees of freedom
diverges as we approach $\Lambda_c$ seems more in consonance with the
picture developed so far. If this is indeed the case, this growth of
the number of degrees of freedom seems to rule out the higher dimensional 
CFT scenario, which would imply a finite number of degrees of freedom at
any energy scale. On the other hand, this behavior has some similarity with 
that of $6d$ little string theories above their mass gap $M^2=M_s^2/N$, where 
a continuum of states develops \refs {\malda, \abks, \ab}, giving (admittedly 
mild) support to the idea that the UV completion ought to be a LST rather 
than a CFT.

In order to decide the existence of such theories and to explore
its properties, it would be important to find a dual realization
of them, in string theory. The first question to address is if
the dual of this RG flow can have a supergravity description in
some regime. For a type II supergravity solution to be reliable, it 
should have small curvature and small dilaton; if the dilaton is large,
a dual description (in 11d SUGRA, or a S-dual solution) would be called 
for. The inverse curvature in string units is typically given in gauge 
variables by $\lambda =g_{YM}^2N$, so near the $\mu \sim
\Lambda _c$ scale, a supergravity dual could have a small curvature. 
Near the IR endpoint, we expect a SUGRA dual description to exist only in
the absence of weakly coupled degrees of freedom on the gauge theory
side, as in the KS cascade.

Concerning the dilaton, the best we can do is to recall
some cases of gravity duals involving non-asymptotically free theories,
and hope that they capture generic features, also present in our duality 
cascades. If we consider the D-brane description of 4+1 and 5+1 SYM with
sixteen supercharges \itzhaki , in both cases at low energies we start
with a description where the effective gauge theory is useful, but being
non-renormalizable, eventually breaks down, being substituted by a D-brane
supergravity description. As the energy scale increases, there is a competing 
effect on the size of the dilaton, between the effective gauge coupling and 
$N$ \itzhaki , and in both cases for large enough energy scale the dilaton 
grows large and we have to resort to a dual description.

For the present duality cascade, the ranks $N$ grow as $N\sim \lambda _+^n$,
with $n$ the number of steps in the cascade, and in the strong coupling 
regime, the coupling grows as $g_i^2\sim (1/r)^n$. Although we have no means 
to determine which particular combination of
gauge couplings and $N$ should enter the expression of the dilaton for a 
potential dual background, given that $1/r$ is larger than $\lambda_+$, the
previous analogy would suggest that eventually such dilaton would grow
large, requiring a dualization of the supergravity description.

In short, for most of the flows, we don't expect a supergravity description 
for the extreme IR of the hyperbolic duality cascade. If we imagine that 
these gauge theories appear in the worldvolume of some brane configuration, 
it is possible that in some regime near $\mu\sim \Lambda _c$ a supergravity 
description is valid, possibly involving a supergravity configuration dual 
to the original one (i.e, in 11d supergravity or the S-dual). The duality
wall might then be realized in supergravity in a similar fashion as the 
mass gap of the LST's is realized in their supergravity duals.

Finally, we could ask how the difference between $\vec N^2>0$ and 
$\vec N^2<0$ flows can show up in the dual. If we assume that some
form of the UV/IR relation holds for the present cascade, since both kinds of
flows have the same UV behavior, we expect them to correspond to string 
backgrounds with the same asymptotic behavior. On the other hand, the 
IR behaviors are quite different, but since we don't expect a supergravity
description to be valid in that regime, it is not clear how this difference
would manifest itself in a dual description.

Needless to say, the present discussion is quite speculative in nature.
We are suggesting the existence of a theory which - given the supposed
grotwh of the number of degrees of freedom - is not a QFT, serving as a 
possible UV completion for the hyperbolic duality cascades discussed in 
this section. One could try to turn this suggestion around, and look for 
4d little string theories (maybe along the lines of \gk ) with a global 
$U(k)\times U(1)$ symmetry, and try to relate them with the duality cascades 
just considered.

\newsec {Conclusions.}

In this work, extending some ideas of \cfikv , we have recasted the 
phenomenon of duality cascade of \ks\ in terms of the Cartan matrix 
underlying the quiver gauge theories that appear in the cascade, and 
we have generalized it to other duality cascades. The Weyl reflections
of the Cartan matrix correspond to Seiberg dualities, and the UV behavior 
of these cascades depends on the kind of Cartan matrix.

One could consider a number of generalizations to the present work.
In \refs{\ahn, \imai, \spcasca} duality cascades involving products
of $SO$ and $Sp$ gauge groups are investigated, and the formalism
presented here could be applied to them. There have been some discussions 
on duality for non 
supersymmetric quiver theories \refs {\schmaltz, \strass, \vafa, \brax}. 
It is clearly important to further study these matters, and to explore
the possibility of non-supersymmetric RG cascades.

We studied in some detail a simple example of gauge theories with
hyperbolic Cartan matrix, and found some evidence for the existence
of some new theory with a fixed scale $\Lambda$ and global
symmetry $U(k)$, which properties somewhat similar to those of little
string theories. Establishing the existence of these theories, and
realizing them by some independent construction, would be quite interesting.

More generally, it is not known which not asymptotically free (NAF) theories
admit a UV completion in terms of an ordinary QFT. In 4 dimensions, 
${\cal N}=4$ gauge theories are always controlled in the UV by the CFT 
at the origin
of moduli space, so the question is trivially answered. For ${\cal N}=2$ 
gauge theories, some results are known: 
$SU(N_c)$ with $N_f$ flavors is NAF for $N_f\geq 2N_c$, but one can argue 
\witten\ that all these theories appear as relevant perturbations of ordinary 
CFT's (although for $N_f$ odd, without a Lagrangian description). Similarly, 
${\cal N}=2$ quiver theories with gauge group $\Pi ~SU(N_i)$ and ADE or affine
ADE Cartan matrix can be obtained by Higgsing of a quiver CFT with the
same Cartan matrix. On the other hand, I am not aware of a straightforward 
UV completion for ${\cal N}=2$ quiver theories with more general Cartan 
matrices, e.g., with hyperbolic Cartan matrices. Although deciding the 
existence or not of such UV completions seems like a hard problem, one might 
hope that progress in this direction would lead towards a new way to deduce 
the existence of new non-gravitational theories, without taking decoupling 
limits from string theory.

\bigskip

\centerline {\bf Acknowledgements.}

I would like to thank O. Aharony and M. Berkooz for many useful 
discusssions regarding the contents of this paper. I have also benefited from 
discussions and correspondence with K. Intriligator, I. Klebanov, 
M. Strassler and A. Uranga. Finally, I would like to thank the organizers of 
the ``Geometric Transitions'' workshop at the CIT-USC Center for Theoretical
Physics for the invitation to attend the workshop, during which part of
this work took place, and for hospitality. This research was supported 
through a European Community Marie Curie Fellowship, and also by the 
Israel-U.S. Binational Science Foundation, the IRF Centers of Excellence 
program, the European RTN network HPRN-CT-2000-00122, and by Minerva.

\vfill
\eject

\listrefs

\end